# Ultra-Thin, High-Lifetime Silicon Nitride Membranes for Nanopore Sensing

Shankar Dutt[1, *], Buddini I. Karawdeniya[2, *], Y.M. Nuwan D.Y. Bandara[1,3], Nahid Afrin[1], and Patrick Kluth[1]

[1]*Department of Materials Physics, Research School of Physics, The Australian National University, Canberra, ACT 2601, Australia.*

[2]*Department of Electronic Materials Engineering, Research School of Physics, The Australian National University, Canberra, ACT 2601, Australia.*

[3]*Research School of Chemistry, The Australian National University, Canberra, ACT 2601, Australia.*

[*]Correspondence: shankar.dutt@anu.edu.au (SD); buddini.karawdeniya@anu.edu.au (BK)

## Abstract

Thin membranes are highly sought-after for nanopore-based single-molecule sensing and fabrication of such membranes becomes challenging in the ≲10 nm thickness regime where a plethora of useful molecule information can be acquired by nanopore sensing. In this work, we present a scalable and controllable method to fabricate silicon nitride ($Si_xN_y$) membranes with effective thickness down to ~1.5 nm using standard silicon processing and chemical etching using hydrofluoric acid (HF). Nanopores were fabricated using the controlled breakdown method with estimated pore diameters down to ~1.8 nm yielding events >500,000 and >1,800,000 from dsDNA and bovine serum albumin (BSA) protein, respectively, demonstrating the high-performance and extended lifetime of the pores fabricated through our membranes. We used two different compositions of $Si_xN_y$ for membrane fabrication (near stoichiometric and silicon-rich $Si_xN_y$) and compared them against commercial membranes. The final thicknesses of the membranes were measured using ellipsometry and were in good agreement with the values calculated from the bulk etch rates and DNA translocation characteristics. The stoichiometry and the density of the membrane layers were characterized with Rutherford backscattering spectrometry while the nanopores were characterized using pH-conductance, conductivity- conductance and power spectral density (PSD) graphs.

**Keywords: Nanopores, Ultra-thin membranes, Controlled breakdown, Nanobiotechnology, DNA sensing, Protein sensing**

## Introduction

Obtaining information from single biomolecules transcends average ensemble approaches and enables methods capable of DNA and protein sequencing. Rapid, portable, low-cost methods and devices for single bio-molecule measurements have gained substantial traction and the current pandemic has recapitulated the need for such devices. Nanopores are often viewed as a modular platform capable of satisfying these criteria with applications spanning, but not limited to, genomics[1], proteomics[2-5], glycomics[6, 7], lipidomics[8], and virology[9, 10]. The broad spectrum of applications of nanopores was greatly enhanced by advancements in nanopore fabrication[11-13], surface decoration[14, 15], material development[16, 17], signal processing algorithms[18, 19], and electronics[20, 21]. A major challenge in nanopore technology is membrane development and pore fabrication, and with the advent of the controlled breakdown technique (CBD)[11], the economics of nanopore fabrication has become more affordable and widespread. A multitude of auxiliary methods has now evolved from CBD expanding the available nanopore fabrication repertoire that can be tailored to user needs and available resources[12, 22-24]. CBD is most suited for membranes that are thinner than ~30 nm. Thinner membranes provide a greater signal-to-noise ratio (SNR), a larger capture radius compared to thicker membranes and enable structural resolution that is a prerequisite for sequencing efforts. With nanopore technology being driven more towards sequencing—may it be genomic or proteomic—fabricating thin membranes has become more desirable for high-resolution measurements.

Silicon nitride ($Si_xN_y$) is the ubiquitous choice of material in most solid-state nanopore (SSN) studies (albeit having high capacitive noise) due to its mechanical, chemical, and thermal stability, availability of thin-film deposition tools, compatibility with silicon-based microelectronics and surface modification approaches. In most studies, the membrane thickness $L_0$ exceeds ~10 nm with some notable examples of $L_0$ ≤5 nm in literature[21, 25-28]. Generally, the fabrication involves methods where a thicker membrane is thinned down to reach a more desired thickness. Typical examples for thinning include laser-thinning[28], electron-beam thinning[29], and ion-beam thinning[13, 30, 31]. Etching methods have mostly been overlooked for the fabrication of thin membranes for SSN studies partly due to the common usage of electron/ion-beam methods. However, unlike the traditional thinning methods, etching methods[32] are scalable, do not require expensive scientific equipment, and deliver wafer-scale fabrication. Owing to the affordability offered by the etching methods, simplifying the thinning process in a reliable and reproducible manner could see widespread access to the coveted <10 nm thickness regime for sensing applications. The ability to tune the $L_0$ in a controllable manner is only a part of the solution since pores fabricated through these membranes should be conducive to analyte translocations. It has been shown previously that a simple change to the electrolyte chemistry during nanopore fabrication could change both nanopore surface chemistry and translocation characteristics[12]. Moreover, surface properties have been shown to influence translocation properties[14, 15, 33].

In this study, we seek to offer a simple, quick, and inexpensive way for creating membranes of any thickness, down to about 3 nm, conducive for biomolecule sensing and characterization applications. Additionally, this study compares the manufactured membranes with commercially available options and investigates the dependence and impact of the membrane film's composition on surface attributes. Thicker membranes (~150 nm and ~200 nm) are first fabricated using microelectromechanical systems (MEMS) processing followed by controlled etching through a series of different concentrations of hydrofluoric acid (HF) for the fabrication of <20 nm thick membranes reaching as low as ~1.5 nm (effective membrane thickness). The stoichiometry, density, and thickness of the fabricated membranes were investigated by Rutherford backscattering spectrometry (RBS) as those parameters can impact the surface charge as well as the pore fabrication characteristics. Typically, amorphous Si-rich $Si_xN_y$ membranes are used for CBD. Here, we present the fabrication of membranes from stoichiometric as well as Si-rich $Si_xN_y$ and the characterization of nanopores in these membranes fabricated using CBD. The nanopores were characterized using pH-conductance (pH-$G_0$) and electrolyte conductivity-$G_0$ measurements to evaluate the pore surface chemistry and surface charge density. Finally, double-stranded DNA (dsDNA) and bovine serum albumin (BSA) were used to explore the suitability of pores for single-molecule sensing. Events >500,000 from dsDNA and as high as ~1.8 million from BSA were achieved from a single solid-state nanopore which by far surpasses the previously reported highest reported number of ~300,000 events. [12]

## Materials and Methods

**Fabrication and Characterization of Membranes:** The workflow for the fabrication of membranes is shown in Figure 1a. As indicated earlier, we fabricated membranes of different compositions as well as with and without a thermal $SiO_2$ underlayer. The workflow for the fabrication is shown for the case of $Si_xN_y$ membranes with thermal $SiO_2$ underlayer. With the slight exception that an LPCVD layer is directly deposited on the wafer, the same fabrication process is used to make membranes devoid of an underlayer. Additionally, because there is no $SiO_2$ underlayer present, there is a slight difference in the etching phases (both reactive ion etching and HF etching). For the fabrication of bare $Si_xN_y$ membranes, ~150 nm $Si_xN_y$ was deposited on both sides of a double-sided polished, 300 μm thick, 4-inch Si wafer employing LPCVD. The LPCVD deposition for bare $Si_xN_y$ was performed at 775 °C and a gas flow of 30 sccm of dichlorosilane and 120 sccm of ammonia was maintained throughout the process to deposit near-stoichiometric $Si_xN_y$ (x~3 and y~4). The thickness of the nitride layer was measured by ellipsometry. Figure 1b shows 61 points of measurement on the wafer along with the measured thickness, indicating almost uniform deposition was achieved with a variation of ~3 nm (standard deviation of 0.95 nm) across the 4-inch wafer. For the case of fabricating $Si_xN_y$ membranes with the $SiO_2$ underlayer, double-sided polished, 300 μm thick wafers with ~100 nm of thermal $SiO_2$ and ~100 nm of low-stress $Si_xN_y$ on both sides were purchased from WaferPro, LLC, US. The $Si_xN_y$ depositions were carried out in a way to achieve near stoichiometric and silicon-rich compositions respectively for bare $Si_xN_y$ membranes and $Si_xN_y$ membranes with thermal $SiO_2$ underlayer. This was done to compare the effect of membrane stoichiometry on the fabricated nanopore surface charge properties. The next steps involve spinning a negative photoresist on the backside of the wafer (iii) and patterning square openings of sizes varying from 430 μm × 430 μm to 550 μm × 550 μm (allowing fabrication of membranes sized 10 μm × 10 μm to 120 μm × 120 μm) using UV lithography (iv). Afterwards, the silicon was exposed from the backside of the wafer in the window area by removing the $Si_xN_y$ layer using reactive ion etching (v). The photoresist was then removed, and the exposed silicon was anisotropically etched by wet etching in 5% tetramethylammonium hydroxide (TMAH, Sigma-Aldrich, 331635) solution at 85 °C (vi). This process leads to the parallel fabrication of 220 membranes of ~200 nm thickness and ~150 nm thickness on a 4-inch Si wafer respectively for the case of fabrication of membranes with and without $SiO_2$ underlayer. While potassium hydroxide has typically been employed as an etchant for anisotropically etching silicon, it also etches silicon oxide at rates up to 10 nm/min[34], which can result in the fabrication of uneven membranes with unknown final thickness. TMAH wet etching offers significantly greater selectivity to etching of $SiO_2$ providing better control over the fabrication process. The material properties of the $Si_xN_y$ layers such as density and stoichiometry were determined by RBS. A 2.0 MeV He ion beam was used to perform RBS on samples and RUMP code[35] was used to simulate and fit the spectra.

**HF etching:** HF etching to thin down the nitride window with and without the silica underlayer was carried out using concentrations of 10%, 5%, and 1% of HF prepared by dilution of 48% HF (Sigma-Aldrich, 695068). The etching was performed in a custom-made etching cradle. To stop the etching, the membranes were rinsed three times in DI water and air-dried.

**Electrolyte preparation:** All electrolytes were prepared by first dissolving the as supplied salts (Sigma-Aldrich, KCl, P9333 and LiCl, L4408) in ~18 MΩ.cm DI water (Sartorius Arium® UV Ultrapure) with the HEPES buffer (Sigma-Aldrich, H0527) followed by filtering through a Millipore Express® PLUS PES filter of 0.22 μm pore size. The pH was tuned to the desired value (±0.1 tolerance) through dropwise addition of either concentrated HCl (Ajax-Finechem, AJA1367, 36%) or KOH (Chem Supply, PA161) and measured with an Orion Star™ pH meter.

**Fabrication of Pores:** The membranes were mounted between two custom fabricated Poly (methyl methacrylate) (PMMA) half cells followed by filling each of the reservoirs with 1 M KCl buffered with 10 mM HEPES at pH ~7. An electric field of <1 V/nm was then applied using a source meter unit (Keithley 2450) until a rapid surge of current was observed which is indicative of pore formation. Afterwards, to characterize the pore size, a current-voltage (I-V) curve was obtained using the Elements eNPR system. The diameter of the pore can be estimated from the slope of the I-V curve (i.e., open-pore conductance, $G_0$) with adequate knowledge of the membrane thickness $L_0$ using the following equation:

$$G_0 = K \left( \frac{1}{\frac{\pi r_0^2}{L_0} + \frac{\mu |\sigma|}{K} \cdot \frac{2\pi r_0}{L_0}} + \frac{2}{\alpha \cdot 2r_0 + \beta \cdot \frac{\mu |\sigma|}{K}} \right)^{-1} \qquad (1)$$

where $G_0$, $r_0$, $K$, $\sigma$, $\mu$, $\alpha$ and $\beta$ are the open pore conductance, nanopore radius, electrolyte conductivity, nanopore surface charge density, mobility of counter-ions proximal to the surface and model-dependent parameters (both $\alpha$ and $\beta$ are set to 2)[36].

**Surface Characterization:** The open-pore conductance was measured as a function of the solution pH using 1 M KCl electrolyte and the data were fitted with eq. 1 using the following approximation for surface charge density ($\sigma$) which then permits the evaluation of the dissociation constant of surface head groups ($pK_a$),

$$|\sigma| \cong \frac{C_{\text{eff}}}{\beta e} W\left(\frac{\beta e}{C_{\text{eff}}} \exp\left((pH - pK_a)\ln(10) + \ln(e\Gamma)\right)\right) \qquad (2)$$

where $e, \Gamma, \beta, C_{eff},$ and $W$ are the elementary charge, number of surface chargeable groups, inverse of the thermal energy, effective Stern layer capacitance, and Lambert $W$ function, respectively[14].

**Biomolecule Sensing:** Double-stranded DNA (Thermo-fisher SM0311) was added to the *cis* chamber to a final concentration of 8.3 ng/µL (~13 nM) and driven across the nanopore in response to a positive bias of 400 mV applied to the *trans* chamber. BSA (Sigma-Aldrich, A7030) was added to the *cis* chamber to a final concentration of 200 nM and driven across the nanopore in response to a positive voltage bias of 500 mV applied to the *trans* chamber. Data were filtered at 10 kHz and sampled at 200 kHz. Collected data were then analyzed using the *EventPro* (3.0) analysis platform[37]. dsDNA and BSA translocation experiments were done using 3.6 M LiCl and 1M KCl buffered at pH ~8 and at pH ~7 respectively.

## Results and Discussion

Membrane fabrication details are outlined in Figure 1a-b and discussed under *Materials and Methods*. Figure 1c shows the RBS spectra recorded along with the fits (solid black line) for both membrane types. The fits to the bare $Si_xN_y$ layer revealed a nearly stoichiometric composition of $Si_3N_{3.94\pm0.02}$ and a density of 2.97 ± 0.02 g cm$^{-3}$. Electron transport is reduced in this material compared to the more commonly used Si-rich $Si_xN_y$ [38]. For the case of silicon-rich membranes with the $SiO_2$ underlayer, the composition and density were found to be $Si_3N_{3.72\pm0.03}$ and 2.94 ± 0.02 g cm$^{-3}$. After the fabrication of thick membranes, they were thinned down in a controlled manner through consecutive etching with HF of different concentrations. We first discuss the case of the near-stoichiometric $Si_3N_{3.94}$ layer. The bulk etch rate calculations using ellipsometry measurements are shown in Figure 1d. Solid lines represent the linear fit to the data revealing the etch rates of 3.78±0.03 nm min$^{-1}$, 1.88±0.04 nm min$^{-1}$, and 0.39±0.03 nm min$^{-1}$ for 10% HF, 5% HF, and 1% HF respectively. The etch rates were also measured using a surface profiler (Bruker Dektak® Stylus Profiler) where half of the sample was covered with a non-etch medium (Polyimide film) before etching. The step height was measured after the etching process and removal of the etch barrier. The thickness of the membrane after different etching times for different concentrations of HF is shown in Figure 1d and the etch rate values obtained from both the surface profiler measurements as well as from the ellipsometry measurements are given in Table S2 for comparison. Values obtained by both methods agree well with each other. With regard to the silicon-rich $Si_3N_{3.72\pm0.03}$ layer, both ellipsometry measurements and surface profiler measurements again revealed comparable bulk etch rates (given in Table S2) for different concentrations of HF. As we need to take account of the underlayer $SiO_2$ while thinning down the $Si_3N_{3.72}$ membranes with $SiO_2$ underlayer, the bulk etch rate for thermal $SiO_2$ was also measured and are indicated in Table S2. The stoichiometric silicon nitride layer exhibits ~25% lower etch rates than the silicon-rich silicon nitride layer, demonstrating the impact of nitrogen concentration on HF etching. The etch rates obtained are almost linear with different concentrations of HF and highly reproducible (within uncertainties) between different wafers.

Figure 1e shows the process flow of thinning down of both membrane types from a starting thickness of ~200 nm ($Si_3N_{3.72}$ with $SiO_2$ underlayer) and ~150 nm ($Si_3N_{3.94}$) membrane to ~5 nm thickness. In brevity, the membranes were etched using 10% HF to a thickness of ~40 nm. As illustrated, the total thickness of the membranes with silica underlayer falls quickly from ~200 nm to ~90 nm within the first 100 seconds of etching owing to the high etch rate of the thermal $SiO_2$ layer as discussed earlier. Then, the membranes were etched with 5% HF to reach a thickness of ~15 nm and finally etched in 1% HF to reach a final thickness of ~5 nm. The small variation of the nitride thickness across the wafer was accounted for during the thinning process by measuring the thickness of the layer right next to the etched window. We fabricated membranes of different sizes using the above-defined method. 220 membranes per 4-inch wafer as small as 10 μm × 10 μm and as large as 120 μm × 120 μm of ~5 nm thickness were fabricated. Window sizes of 40 μm × 40 μm were chosen for the experiments reported in this study. Although thinning of silicon nitride is also possible by the reactive ion etching method, due to the intrusive nature of the plasma caused by the bombardment of the surface with ions as well as due to the creation of defects[39], membranes thinner than ~25 nm did not survive. On the other hand, using different concentrations of HF with well-characterized etch rates, we created membranes as thin as ~3 nm thickness and demonstrated the excellent stability of the fabricated membrane for nanopore fabrication and biosensing.

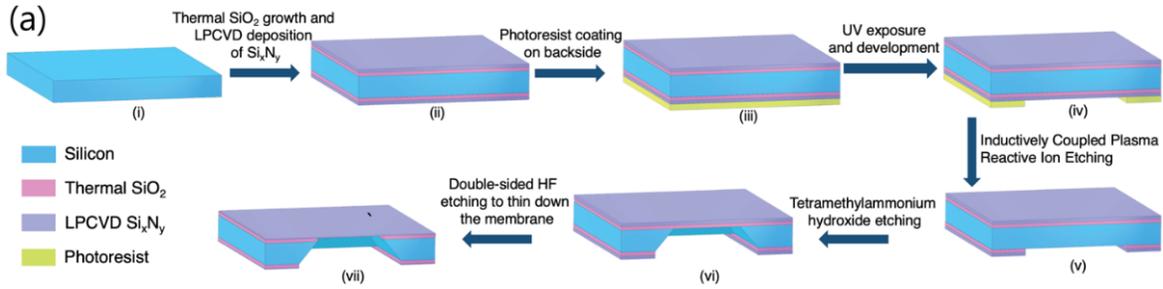

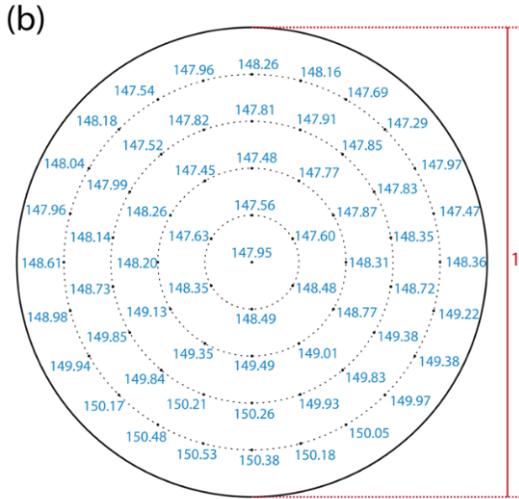
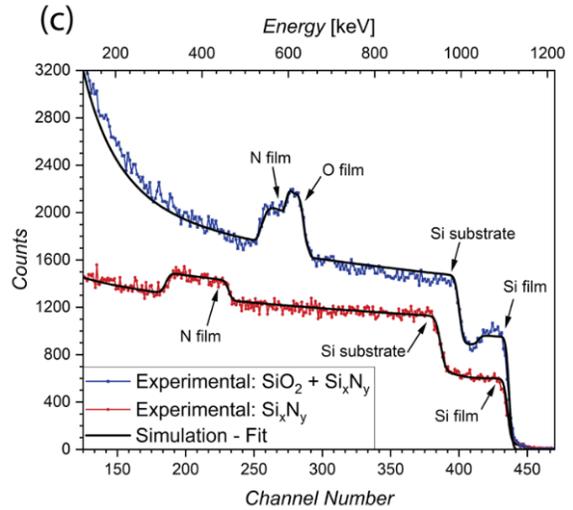

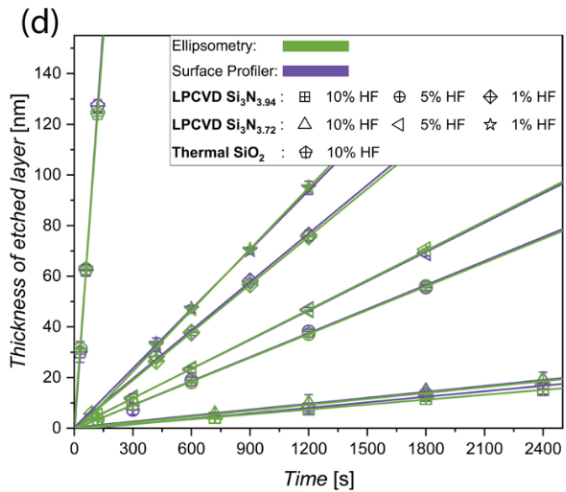
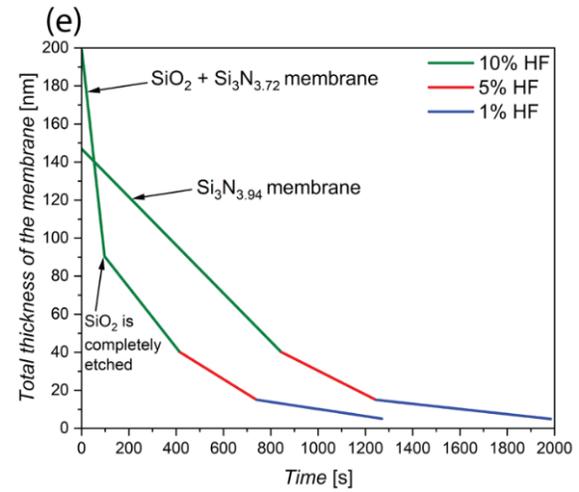

**Figure 1: (a)** Process flow showing the steps for the fabrication of ultrathin silicon nitride membranes with thermal SiO$_2$ underlayer. **(b)** 61 different points measured by ellipsometry on a 4'' wafer with thicknesses of the deposited bare silicon nitride film in nm. **(c)** Rutherford backscattering spectra of the bare stoichiometric Si$_x$N$_y$ (red) and silicon-rich Si$_x$N$_y$ (blue) with thermal SiO$_2$ underlayer along with the fits using the RUMP code (black solid line). **(d)** Thickness of the etched layer (for the case of Si$_3$N$_{3.94}$, Si$_3$N$_{3.72}$ and thermal SiO$_2$ layers) as a function of etching time for different concentrations of HF as measured by ellipsometry and surface profilometry. The solid lines represent the linear fit to the data depicting the etch rates for different concentrations. **(e)** Process flow for reducing the thickness of the membranes (bare $Si_3N_{3.94}$ and $Si_3N_{3.72}$ with $SiO_2$ underlayer) as a function of time using different concentrations of HF to obtain good control over the process. The process is controllable and can be stopped at different points to obtain membranes with desired thickness.

After the controlled thinning of the membranes, pores were fabricated using the CBD method as shown in Figures 2a and 2b (see *Fabrication of Pores* under *Materials and Methods* for more details). For near-stoichiometric $Si_xN_y$ (i.e., $Si_3N_{3.94}$) membranes, it typically takes ~2-3 minutes and ~25-30 minutes for the initial breakdown to take place for ~5 nm and ~12 nm thick membranes, when 3-3.5 V and 7-7.5 V is applied across the membrane, respectively. Compared to Si-rich $Si_xN_y$, (i.e., $Si_3N_{3.72}$) this is about double the time required, which is not surprising given the near stoichiometric nature of the membranes used for this study. The slow breakdown facilitated the precise fabrication of <5 nm diameter pores. Typical I-V curves of the pores are shown in Figure 2c (through near-stoichiometric $Si_xN_y$ membranes) which instantaneously exhibited Ohmic behavior without requiring overnight soaking or any pretreatment. I-V curves from pores fabricated through Si-rich $Si_xN_y$ membranes were instantaneously Ohmic as well (data not shown). This is especially advantageous since the fabricated pores could be directly used for sensing applications. The pores depicted in Figure 2c ranged from ~1.8 nm (6 nS) to ~5.6 nm (35 nS) in diameter (determined using equation (1)). Figure 2d shows the thickness of 11 near-stoichiometric $Si_xN_y$ membranes. The red ribbon in the figures depicts the target thickness (~5.0±0.5 nm). We see that all membranes were <8 nm in thickness with most membranes falling well within the expected thickness bracket with around ±2.5 nm deviation, which is typical for commercial membranes as well. Similarly, Figure 2e shows the thickness of 11 Si-rich $Si_xN_y$ (i.e., $Si_3N_{3.72}$) membranes with a target thickness of ~5.0±0.5 nm. All membranes were <8 nm in thickness and less scattered compared to their stoichiometric counterpart. This demonstrates the application of our method for fabrication of thin membranes in a controllable manner irrespective of stoichiometry. The thickness control of the membranes can be further improved by more precise control of the HF concentration and etching temperature.

Next, we looked at the surface chemical properties of the nanopores. The nanopore chemistry influences the translocation properties and can be tuned to slow down analyte transport[14, 15, 40] and selectively capture analytes[41]. The surface chemistry is also inextricably linked to properties such as the capture rate, open-pore stability, transient or incessant surface-analyte interactions, transport mechanism, and signal-to-noise ratio (SNR), to name a few. Although probing the inner nanopore surface is challenging due to the constricted volume, pH-conductance (pH-$G_0$) and electrolyte-conductance (K-$G_0$) can reveal significant information about the nature of surface head groups (e.g., $pK_a$) and surface charge (e.g., $\sigma$) of the inner nanopore surface. To this extent, $G_0$ was evaluated as a function of pH as shown in Figures 2f for near-stoichiometric $Si_xN_y$ and Figure 2g for Si-rich $Si_xN_y$. The raw data were fitted with equations 1 and 2 which yielded a pKa of 8.0±0.4 (from 5 unique pores) for near-stoichiometric $Si_xN_y$. This pKa value suggests that the surface is rich with acidic head groups and is in good agreement with that reported for silanol groups which would imply that the surface hydroxyl groups are responsible for the observed value[42]. In contrast, Si-rich non-stoichiometric $Si_xN_y$ exhibits an amphoteric behavior with an isoelectric point (pI) ~4.0±0.5 (from 3 unique pores, Figure 2h). The pH-$G_0$ curves of the Si-rich $Si_xN_y$ commercial membranes with and without the $SiO_2$ underlayer resembled Figure 2h with a pI of ~3.7 and ~4.3 respectively (Figure S1). Thus, we see a close agreement between the pI values of our membranes and the commercial ones with both categories in good agreement with the previously reported values[14, 43]. Recent work demonstrating pore fabrication in non-stoichiometric $Si_xN_y$ using chemically tuned controlled dielectric breakdown (CT-CDB)[12] and Tesla coil assisted method (TCAM)[22] showcased the role of electrolyte chemistry during fabrication where pH-$G_0$ curves resembled Figure 2f rather than Figure 2g. These results in conjunction with ours show that the inner nanopore surface chemistry depends on a host of factors such as membrane stoichiometry, fabrication method and electrolyte chemistry. Afterwards, $G_0$ was evaluated as a function of electrolyte conductivity (pH ~8) as shown in Figures 2h and 2i. The observed pattern is typical and the deviation from the linear behavior at low electrolyte concentrations has been attributed to the increasing contribution from the nanopore surface charge to the overall conductance. The dashed lines in both the figures are fits to the raw data using equation 1 where $L_0$ was constrained to a maximum of 10 nm. This yielded a $\sigma$ of ~4.6±1.0 mC m$^{-2}$ (2 unique pores) and ~5.6±1.7 mC m$^{-2}$ (3 unique pores) for pores fabricated through near-stoichiometric $Si_xN_y$ and Si-rich $Si_xN_y$ membranes, respectively. Although the averaged $\sigma$ value for the Si-rich $Si_xN_y$ is higher than that of the near-stoichiometric $Si_xN_y$, statistically they are within error. Since $\sigma$ at this pH value (i.e., ~8) predominantly arises from the dissociative equilibrium of acidic head groups which is silanol in both instances, unless the density of surface functional groups is different in the two cases, $\sigma$ is expected to be in close agreement.

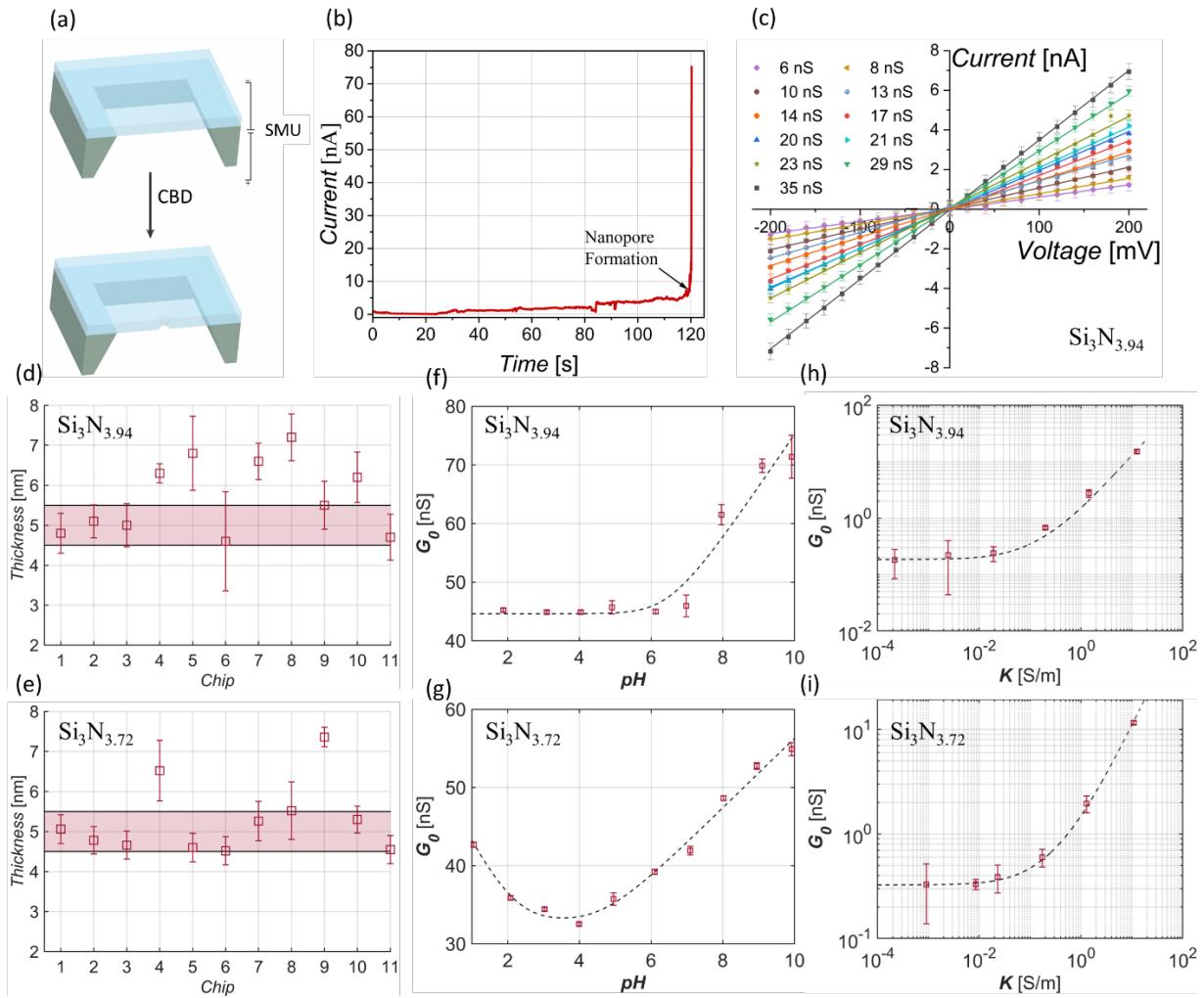

**Figure 2: (a)** Schematic of nanopore fabrication by CBD using a source meter unit (SMU) where a voltage is applied until **(b)** a sudden rise in current is seen (in 1M KCl). **(c)** I-V curves for a range of pore sizes fabricated using the CBD method. They instantaneously showed Ohmic behavior, and the size ranged from ~ 1.8 nm (6nS) to ~ 5.6 nm (35 nS) (see Table S1 for more details). Sizes were estimated using equation 1. SI Table S1 shows the complete set of pore diameters for the I-V curves shown in (c). The thickness of 11 representative **(d)** near stoichiometric $Si_3N_{3.94}$ chips (See SI table S1 for more details of these chips) and **(e)** $Si_3N_{3.72}$ chips with $SiO_2$ underlayer that were used for pore fabrication. The target thickness is represented by the red band (~5.0±0.5 nm). Measurement of open-pore conductance ($G_0$) with pH of **(f)** near stoichiometric $Si_3N_{3.94}$ and **(g)** $Si_3N_{3.72}$ chips with $SiO_2$ underlayer. Measurement of open-pore

conductance ($G_0$) with the electrolyte conductivity (KCl buffered at pH ~8) of **(h)** near stoichiometric Si$_3$N$_{3.94}$ chips and **(i)** Si$_3$N$_{3.72}$ chips with SiO$_2$ underlayer with the fits corresponding to equation 1.

We compared the noise properties of the membranes fabricated in this study against the commercial ones. As seen in Figure 3a, 4 membrane types were used for comparison: 2 in-house and 2 commercial ones with and without SiO$_2$ underlayer. The thickness of the membranes used for Figure 3a were 10.7±1.1 nm (Si$_3$N$_{3.94}$, without SiO$_2$ underlayer), ~12 nm (Norcada, without SiO$_2$ underlayer) 11.1±0.8 nm, (Si$_3$N$_{3.72}$, with SiO$_2$ underlayer) and ~12 nm (Norcada, with SiO$_2$ underlayer). The thickness of the membranes fabricated using our fabrication method was kept comparable to that of the commercial membranes so that a direct comparison can be made. The pore diameters were ~3.2 nm, ~3.1 nm, ~3.6 nm, and ~3.4 nm, respectively, for the 4 membrane types. The SiO$_2$ underlayer has been previously shown to reduce the dielectric noise.[44] The dielectric noise ($S_d$) is given by $S_d = 8\pi kTCDf$ where $k, T, C, D$ and $f$ are the Boltzmann constant, temperature, non-ideal capacitance, dielectric loss factor and frequency. Thus, a material with a lower $D$ will have a lower $S_d$. We see that the standard deviation ($std$) of the current trace reduces by a factor of approximately 2 with the introduction of the SiO$_2$ underlayer with both in-house and commercial membranes. More interestingly, we see that the $std$ values of the in-house and commercial substrates are comparable and mostly within ~10% of each. Considering the customizability associated with the in-house method and the ability to fabricate <10 nm thick membranes, we see this as a significant improvement for the development of recipes for membrane fabrication aimed towards nanopore sensing. The power spectral density (PSD) curves (Figure 3b) further corroborate the observations in Figure 3a where we see the membranes with the SiO$_2$ layer showing lower noise (green and red curves) compared to the SiO$_2$-free ones (blue and orange curves).

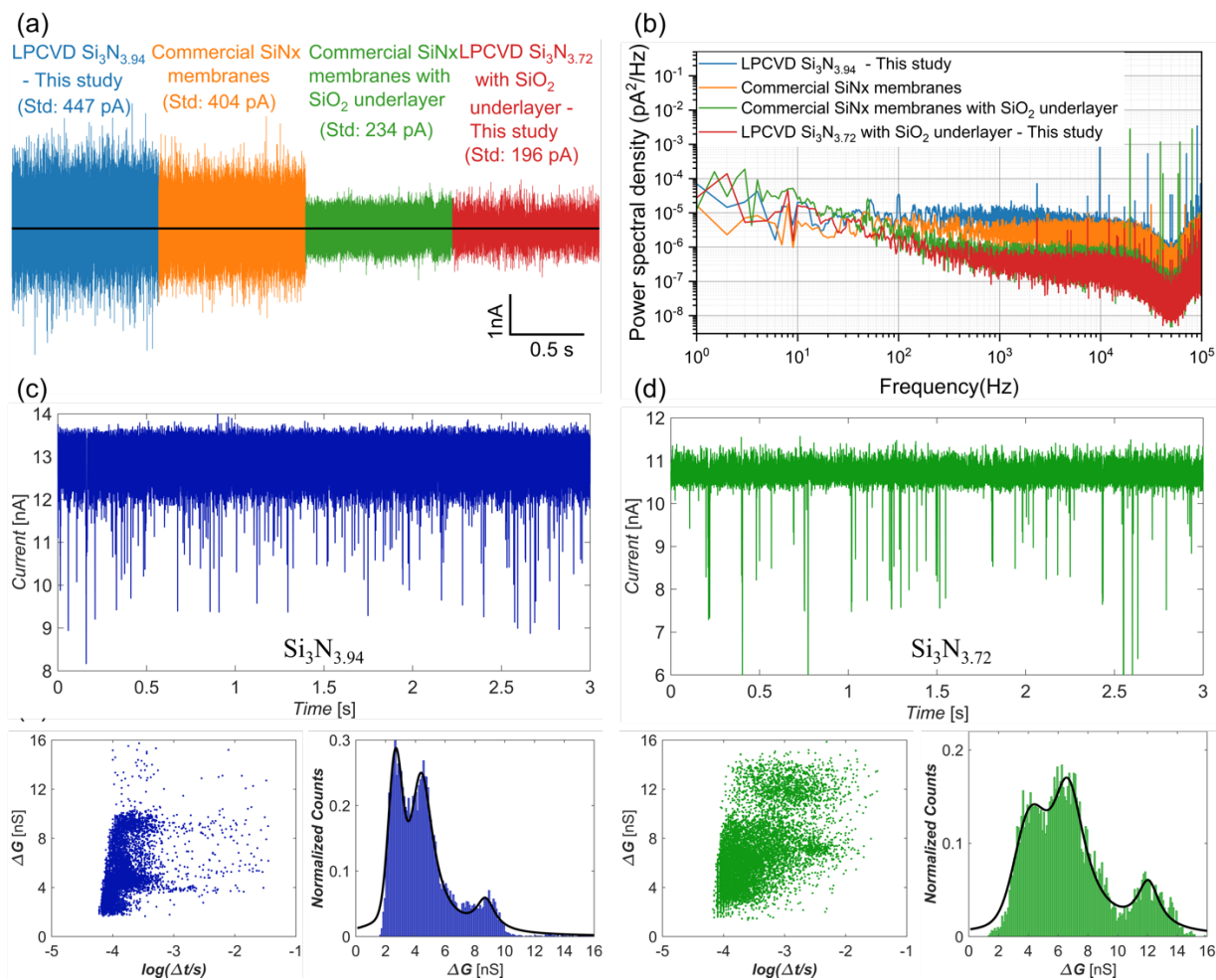

**Figure 3**: Representative 1-second open pore current traces corresponding to membrane types investigated in this study. The standard deviation of the current trace as a metric of noise is indicated in parenthesis. **(b)** The power spectral density (PSD) curves of the traces depicted in (a). **(c)** and **(d)** Representative current traces, scatter plots and histograms (corresponding to change in conductance (ΔG)) resulting from the translocation of dsDNA through ~9.8 nm (blue) ~6.4 nm (green) thick near-stoichiometric and non-stoichiometric $Si_xN_y$ membranes, respectively. The thickness was calculated using the *molecular capillary* method shown in Figure S2 (and the discussion under *Molecular Capillary Method* in the SI)[45, 46]. The pore diameters were ~5.4 nm and ~5.2 nm respectively. All experiments were conducted with 8.3 ng/μL dsDNA in 3.6 M LiCl buffered at pH ~8 with a bias of +400 mV.

After the fabrication of pores in thin membranes and the chemical characterization of the nanopore surface, we conducted translocation experiments using dsDNA and BSA. The purpose of using dsDNA in this study is twofold: (i) to show that the fabricated pores through the thinned down membranes are conducive for analyte translocation and (ii) to corroborate the membrane thickness from methods discussed previously using dsDNA translocation characteristics.[45, 47] We used 3.6 M LiCl buffered at pH ~8 for the dsDNA translocation experiments. Under such high electrolyte concentrations, electroosmosis would be limited. We opted for LiCl instead of KCl since it is known to slow down the translocation of DNA.[48] Representative current traces from two different pores for each of the in-house fabricated membranes are shown in Figures 3c and 3d. Histograms corresponding to the change in conductance due to dsDNA translocation ($\Delta G$) were fitted with a Lorentzian-Gaussian mixture model.[37] The first $\Delta G$ distribution ($\Delta G_0$) is often attributed to collisions while the other two (in the order of increasing $\Delta G$) are attributed to single file ($\Delta G_1$) and folded over ($\Delta G_2$) conformations of dsDNA. For further analysis, the single file and folded-over translocations were separated using Gaussian Mixture clustering (see SI figure S3 for more information).[37] For dsDNA, the ratio of the third and second peak (i.e., $\Delta G_2/\Delta G_1$) is typically ~2. From the fits corresponding to $\Delta G_1$ and $\Delta G_2$ (SI figure S1), the ratios $\Delta G_2/\Delta G_1$ corresponding to pores through stoichiometric ($Si_3N_{3.94}$) and non-stoichiometric ($Si_3N_{3.72}$) membranes were computed to be ~1.94 and ~1.96, respectively, which is in close agreement with the ideal value of ~2 for dsDNA. Here we note, without the clustering of the single file and folded-over conformations, the ratio $\Delta G_2/\Delta G_1$ from the fits shown in Figure 3c and 3d corresponding to pores through stoichiometric ($Si_3N_{3.94}$) and non-stoichiometric ($Si_3N_{3.72}$) membranes yielded ~1.82 and ~1.99 which is still in good agreement with the ideal value of ~2. The change in conductance because of dsDNA passage ($\Delta G_{dsDNA}$) was then used as an independent metric to estimate $L_0$ (i.e., *molecular capillary* method, see Figure S2 and the discussion under *Molecular Capillary Method* for more details) from which the $L_0$ were found to be ~9.8 nm (near-stoichiometric membrane), and ~6.4 nm (non-stoichiometric membrane). The corresponding values estimated from the ellipsometry measurements were $8.8 \pm 0.9$ nm, and $7.2 \pm 1.1$ nm, respectively, which are in good agreement with the values calculated using the *Molecular Capillary Method*.

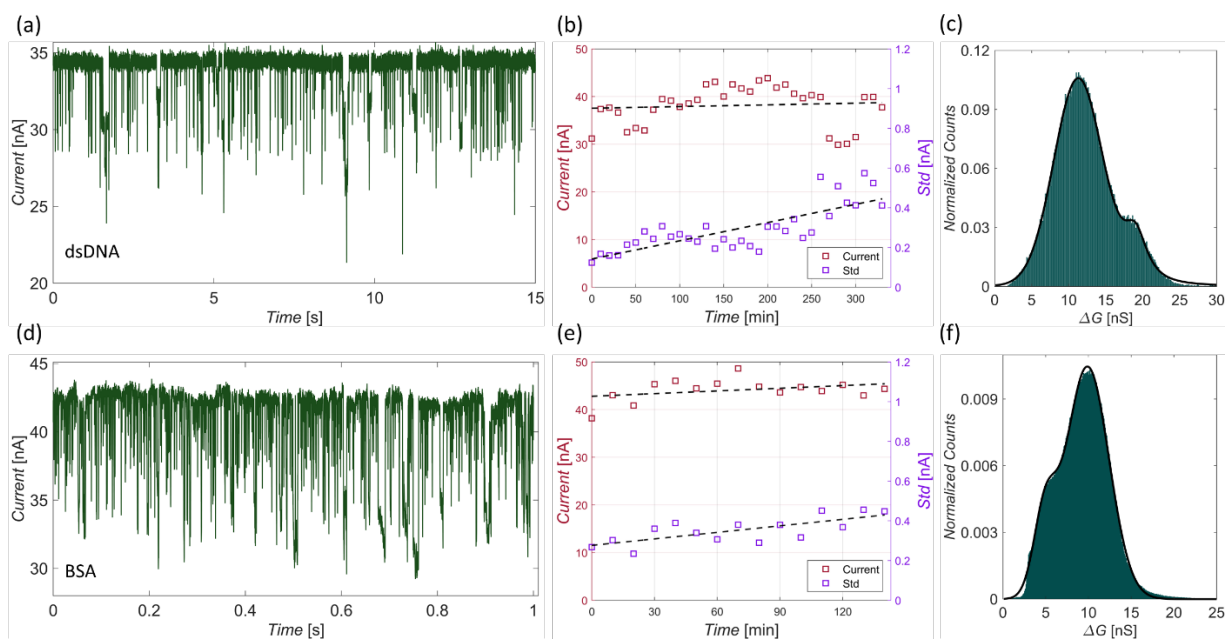

**Figure 4: (a)** 15-second current trace from a ~4.8 nm diameter pore through a ~1.5 nm thick $Si_3N_{3.72}$ membrane under +400 mV of applied voltage. **(b)** Progression of open-pore current and its standard deviation with time. The pore yielded a total of 517,993 events and the histogram corresponding to ΔG is shown in **(c)**. **(d)** 1-second current trace from a ~14.1 nm diameter pore through a ~5.4 nm thick $Si_3N_{3.72}$ membrane under +500 mV of applied voltage. **(e)** Progression of open-pore current and its standard deviation with time. The pore yielded a total of 1,832,151 events and the histogram corresponding to ΔG is shown in **(f)**. The dashed lines in (b) and (e) are linear fits made to the raw data.

The thinning method was then extended to fabricate the thinnest Si-rich membrane (i.e., $Si_3N_{3.72}$) of this study: ~4.8 nm diameter pore through a ~1.5 nm (effective thickness) thick membrane was fabricated using the CBD method. The pore was run for >5 hrs from which we were able to acquire >500,000 events (517,993 events). This is, to the best of our knowledge, the highest event count reported from a single solid-state silicon nitride nanopore which further demonstrates the significant advancement we have made in membrane technology conducive to dsDNA translocation. The membrane thickness estimated from the ellipsometry measurement was 3.1±0.8 nm. We note that the effective thickness determined through the *Molecular Capillary Method* is half the measured thickness. Such discrepancy is not uncommon[45] and we attribute it to deviations from the cylindrical geometry in the pore opening which can be significant at these thickness scales. The local thickness around the nanopore may thus be thinner than the membrane and this is what is reflected in the *molecular capillary* calculations. Figure 4a shows a 15-second trace of this pore while Figure 4b shows its temporal stability over time. We see that both the open-pore current and noise (standard deviation of the open-pore current) increase over time, which is not unusual for solid-state nanopores. Through a linear fit, the rate of increase was found to be ~$3.5 \times 10^{-3}$ nA/min and ~$9 \times 10^{-4}$ nA/min for open-pore current and its noise, respectively. Moreover, to see a growth of ~0.5 nm in the pore diameter (~10% change of the initial pore diameter), it would take ~9 hours of continuous operation of the pore provided the growth kinetics stay uniform throughout. Furthermore, for a 10% increase in the baseline noise, it would take about ~2.5 hrs. of continuous operation. Thus, we see that the pore noise increase ~3.6× faster compared to the pore diameter. This is somewhat expected since prolong translocation of an analyte through a pore often leads to an increase in the baseline noise. However, even after running the pore for ~5 hrs., the increase in the pore-noise was insignificant to interfere with the pore-quality needed for event extractions partly due to deeper event blockades observed with the translocation experiments (Figure 4a). We observed signs of intermittent analyte clogging appearing with continuous pore-operation which is evident through the abrupt drop in open-pore current after 4 hours of operation (Figure 4b). However, the pore could be easily revived by the application of a *zap-potential* (a common practice in the field). Furthermore, the histogram corresponding to ΔG shows 2 peaks (Figure 4c) unlike those presented in Figure 3 indicating that the *bumping* events are near-absent with the thinner membranes. Moreover, the signatory dual peak distribution is still visible, yet, not as separated as those in Figure 3c and 3d. While the pore diameters in the two cases are not far apart, the results hint that single-file translocations are more favored with thinner membranes. This is a crucial finding since sequencing efforts prefer the minimization of 3D conformational distribution in the translocation population.

Protein sensing can be challenging for a multitude of reasons and more aggravated by their propensity to (non-specifically) stick to nanopore surfaces and thereby occluding the ion passage irreversibly. We chose one of the stickiest proteins, BSA, which is well known to nonspecifically bind to a host of surfaces. Furthermore, to test the temporal stability of the pore and its resilience against clogging by the protein, we used a high concentration (200 nM) of BSA for the translocation experiments. For the experiment shown in Figure 4d, we used a ~14.1 nm diameter pore through a ~5.4 nm thick membrane (estimated through ellipsometry) where BSA travelled electrophoretically across the nanopore generating resistive pulses. We were able to collect over 1.8 million (1,832,151) events with this pore while three more pores (~14.1 nm, ~14.7 nm, and ~16.3 nm in diameter) yielded 625,033, 441,396, and 944,605 events, respectively. Moreover, similar to Figure 4b, the open-pore current and noise (standard deviation of the open-pore current) were calculated for the pore shown in Figure 4c which stayed operational for ~3 hours (Figure 4e). We observe that the open-pore current and noise increase with time at a rate of ~$18.7 \times 10^{-3}$ nA/min and ~$11 \times 10^{-4}$ nA/min respectively.

While the growth kinetics of the pore used for DNA experiments are better than that used for the BSA experiments, in the context of the time it takes for a 10% change in the pore diameter and noise (~6 hrs. and ~4.5 hrs. respectively), the pores indicate great stability for lengthy experiments. Thus, we see that a significant change in the pore diameter or noise does not occur even after running the experiments for few hours. The ΔG distribution shows a bimodal distribution where the first peak appears to be a shoulder of the major peak centered at ~10 nS. We believe these originate from the different entrance trajectories and entrance shapes of the prolate BSA.[49] The DNA and BSA results taken together demonstrate the exceptional stability of the pores fabricated in our thin membranes and their suitability for long duration nanopore sensing efforts.

## Conclusion

In this study, we have demonstrated a controllable method to fabricate thin $Si_xN_y$ membranes on wafer-scale. Moreover, membranes of <10 nm thickness can be fabricated conveniently with the ability to reach thicknesses as low as ~1.5 nm (effective thickness). Both stoichiometric and Si-rich silicon nitride membranes were fabricated, and their chemistry was confirmed through RBS. Single nanopores were then formed using the CBD method and the surfaces of the fabricated pores were probed using pH-$G_0$ and K-$G_0$ curves. From the former, the nanopore surfaces through stoichiometric membranes were found to be rich in acidic head groups while those formed from non-stoichiometric (Si-rich) membranes were amphoteric in nature. The pKa for pores fabricated through stoichiometric membranes was found to be in close agreement with that of a silanol-rich surface. The isoelectric point of pores fabricated through Si-rich membranes was also in good agreement with reported values. The change in open-pore conductance for dsDNA translocations were measured with a range of pores which produced translocation characteristics commensurate with the structure of DNA and dimensions of the pore. The $L_0$ calculated from dsDNA translocations further corroborated the thickness predicted by the surface profiler and ellipsometry techniques. The noise levels were compared with similar commercial products where the results demonstrated comparable performance. Furthermore, with dsDNA, we were able to collect events as high as >500,000—the highest reported from a single silicon nitride nanopore to the best of our knowledge. On the protein front, we were able to supersede this limit by collecting over ~1.8 million events. This, to the best of our knowledge is the first time where the 1 million event barrier was breached by a single solid-state nanopore. Our findings would be beneficial for the widespread adoption of nanopore technology as it presents a convenient and scalable membrane fabrication method conducive for nanopore fabrication by CBD and analyte translocations.

## Acknowledgments

This work used the ACT node of the NCRIS-enabled Australian National Fabrication Facility (ANFF-ACT). BIK, PK and YMNDYB were supported by the 'Our Health in Our Hands' ANU Grand Challenge. SD was supported by an AINSE Ltd. Postgraduate Research Award (PGRA) and the Australian Government Research Training Program (RTP) Scholarship. PK also acknowledges financial support from the Australian Research Council (ARC) under the ARC Discovery Project Scheme (DP180100068). Authors acknowledge access to NCRIS facilities (ANFF and the Heavy Ion Accelerator Capability) at the Australian National University.

## Disclaimers

None

## Data Availability

Data will be available upon reasonable request.

# Ultra-Thin, High-Lifetime Silicon Nitride Membranes for Nanopore Sensing

Shankar Dutt[1], Buddini I. Karawdeniya[2], Y.M. Nuwan D.Y. Bandara[1,3], Nahid Afrin[1], and Patrick Kluth[1]

[1] Department of Materials Physics, Research School of Physics, The Australian National University, Canberra, ACT 2601, Australia.

[2] Department of Electronic Materials Engineering, Research School of Physics, The Australian National University, Canberra, ACT 2601, Australia

[3] Research School of Chemistry, The Australian National University, Canberra, ACT 2601, Australia.

**Table S1:** Membrane thickness (estimated from the surface profiler, with 3× the relative error), $G_0$ (from I-V curves in Figure 2c), estimated pore diameter (using equation 1). All I-V curves were measured using 1M KCl (K~12.69 S/m) buffered at pH ~7.

| Membrane thickness (nm) | $G_0$ (nS) | Calculated Pore Diameter |
|---|---|---|
| 4.8 ± 0.5 | 5.8 ± 0.1 | 1.8 ± 0.1 |
| 5.1 ± 0.4 | 7.7 ± 0.1 | 2.3 ± 0.1 |
| 5.0 ± 0.5 | 10.4 ± 0.1 | 2.7 ± 0.2 |
| 6.3 ± 0.2 | 13.0 ± 0.3 | 3.4 ± 0.1 |
| 6.8 ± 0.9 | 14.3 ± 0.1 | 3.7 ± 0.5 |
| 4.6 ± 1.2 | 17.4 ± 0.2 | 3.5 ± 0.6 |
| 6.6 ± 0.5 | 19.7 ± 0.1 | 4.4 ± 0.3 |
| 7.2 ± 0.6 | 20.8 ± 0.1 | 4.7 ± 0.4 |
| 5.5 ± 0.6 | 23.0 ± 0.2 | 4.5 ± 0.3 |
| 6.2 ± 0.6 | 28.8 ± 0.1 | 5.4 ± 0.4 |
| 4.7 ± 0.6 | 35.1 ± 0.1 | 5.6 ± 0.3 |

**Table S2:** Etch rates (in nm min$^{-1}$) of different layers estimated by ellipsometry and surface profiler in 10%, 5% and 1% HF.

|  |  | Etch rate in 10% HF (nm min$^{-1}$) | Etch rate in 5% HF (nm min$^{-1}$) | Etch rate in 1% HF (nm min$^{-1}$) |
|---|---|---|---|---|
| Si$_3$N$_{3.94}$ | ellipsometry | 3.78±0.03 | 1.88±0.04 | 0.39±0.03 |
| | surface profiler | 3.82±0.14 | 1.89±0.08 | 0.43±0.04 |
| Si$_3$N$_{3.723}$ | ellipsometry | 4.78±0.04 | 2.33±0.02 | 0.46±0.03 |
| | surface profiler | 4.68±0.08 | 2.31±0.03 | 0.47±0.03 |
| SiO$_2$ | ellipsometry | 61.91±0.35 | - | - |
| | surface profiler | 64.76±1.51 | - | - |

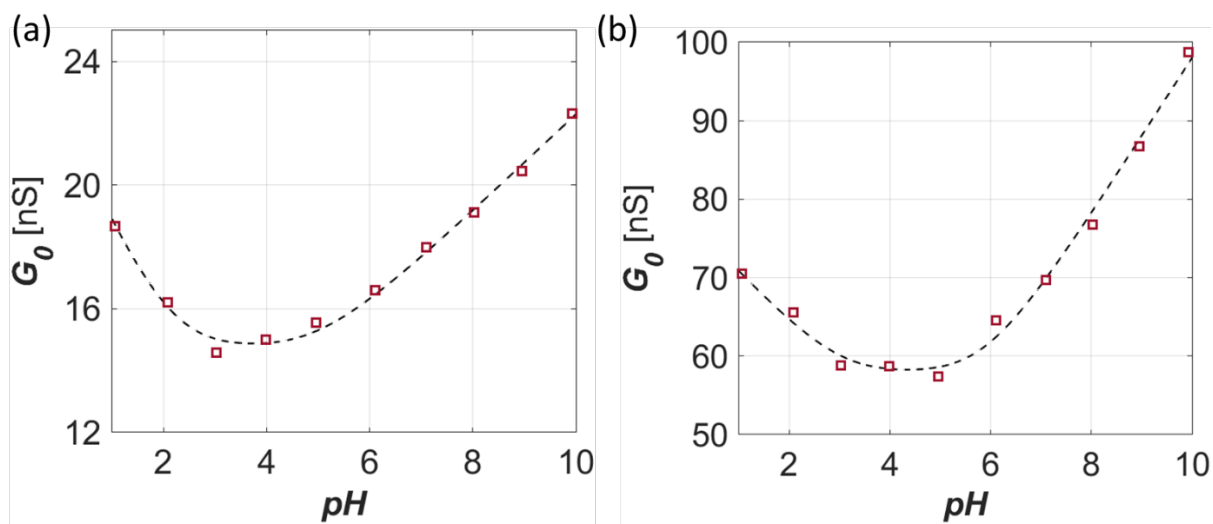

**Figure S1:** pH-$G_0$ curves corresponding to commercial membranes (Norcada) **(a)** with **(b)** and without the SiO$_2$ underlayer. All experiments were done with 1M KCl.

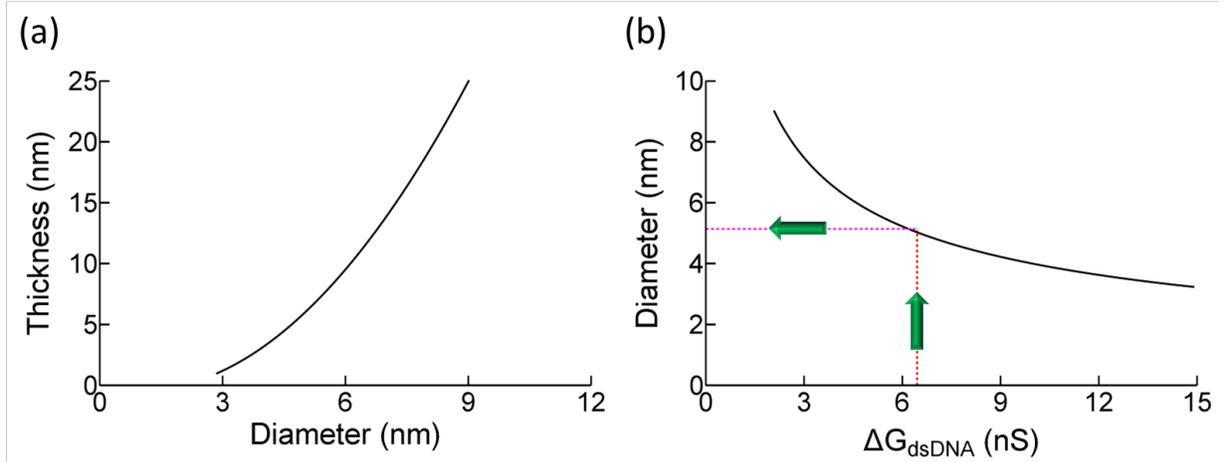

**Figure S2:** (**a**) The relationship between membrane thickness and pore diameter for a pore with a conductance ($G_0$) of ~20.8 nS (electrolyte conductivity ~10.47 S/m) calculated using equation 1. Without the knowledge of pore diameter or membrane thickness, there are infinite solutions to pore diameter and membrane thickness that would satisfy $G_0$. (**b**) The relationship between pore diameter $\Delta G_{dsDNA}$. The red dashed line shows $\Delta G_1 \sim$ 6.2 nS (single file translocations of dsDNA) for the translocations shown in Figure 3d. The calculated values through the intersection point were ~5.2 nm and ~6.4 nm for diameter and membrane thickness, respectively.

**Molecular Capillary Method**

The change in conductance because of dsDNA passage ($\Delta G_{dsDNA}$) was used as an independent metric to estimate $L_0$ (i.e., *molecular capillary* method). $\Delta G_{dsDNA}$ can be modeled using, $\Delta G_{dsDNA} = G_0 - \left( \dfrac{1}{K \frac{\pi r_{0,with\ dsDNA}^2}{L_0}} + \dfrac{2}{G_{acc,DNA}} \right)^{-1}$ where $r_{dsDNA}$, $r_{0,with\ dsDNA}$ and $G_{acc,DNA}$ are the radius of dsDNA (set to 1.1 nm), open-pore radius when dsDNA is inside the pore, $\left( \sqrt{r_0^2 - r_{dsDNA}^2} \right)$ and access conductance in the presence of dsDNA $\left( 4Kr_0 - \dfrac{K\pi(2r_0)^2}{4r_0} \right)$, respectively.[1] For this, we disregarded the surface contributions, both from the nanopore surface and dsDNA due to the high salt concentration in the experiments. Thus, $G_0$ of equation 1 reduces to $K\left(\frac{1}{\frac{\pi r_0^2}{L_0}} + \frac{2}{\alpha \cdot 2r_0}\right)^{-1}$ (see equation 1 for the full form). Thus, the knowledge of $\Delta G_{dsDNA}$ can be used to find $L_0$ (or $r_0$) and by extension $r_0$ (or $L_0$) as shown in SI Figure S2b.

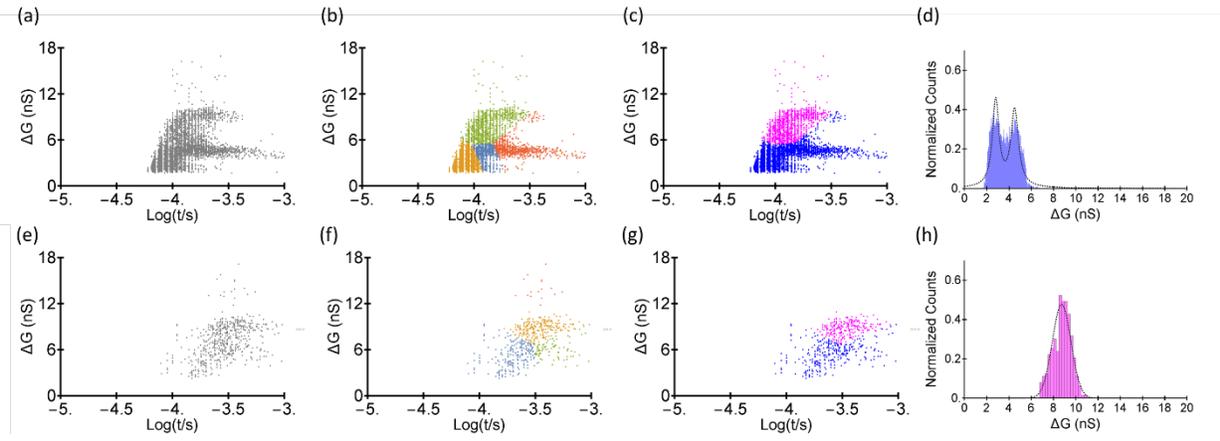

**Figure S3:** (a) single-level events corresponding to ~9.4 nm thick stoichiometric membrane of Figure 3c. Two populations can be seen since not only single files but also looped conformations are detected as single-level events. (b) The scatter plot is then subjected to clustering using the "*Gaussian-Mixture*" method of Mathematica. (c) Populations with similar ΔG are combined to get two populations with the lower one corresponding to single file translocations (blue) and the upper one signifying looped conformations (magenta). (d) The histograms corresponding to single file translocations showed two populations with the lower ΔG population assigned to collisions and the higher ΔG to true translocations. The histogram was then fitted using a Gaussian Mixture Model to extract $\Delta G_1$. (e) multi-level events corresponding to ~9.4 nm thick stoichiometric membrane of Figure 3c. Two populations can be seen since events with shallow

steps are grouped into the multi-level category.[2] (**f**) The scatter plot is then subjected to clustering using the "*Gaussian-Mixture*" method of Mathematica. (**g**) Populations with similar ΔG are combined to get two populations with the lower one corresponding to events with shallow steps (blue) and the upper one signifying folded-over conformations (magenta). (**h**) The histograms corresponding to folded-over translocations showed a single population and was fitted with Gaussian model extract $\Delta G_2$.